# A search for mid-IR bands of amino acids in the Perseus Molecular Cloud

Susana Iglesias-Groth

*Instituto de Astrofísica de Canarias,* C/ Vía Láctea s/n, 38200 La Laguna, Tenerife, Spain

**Abstract**

**Amino acids are building-blocks of proteins, basic constituents of all organisms and essential to life on Earth. They are present in carbonaceous chondrite meteorites and comets, but their origin is still unknown. Formation of amino acids in the interstellar medium is posible via specific gas-phase reactions in dark clouds, however sensitive radiosearches at millimeter wavelengths have not revealed their existence yet. The mid-IR vibrational spectra of amino acids provides an alternative path for their identification. We present Spitzer spectroscopic observations in the star-forming region IC 348 of the Perseus Molecular Cloud showing the detection of mid-IR emission lines consistent with the most intense laboratory bands of the three aromatic amino acids, tyrosine, phenylalanine and tryptophan and the aliphatic amino acids isoleucine and glycine. Estimates of column densities give values 10-100 times higher for isoleucine and glycine than for the aromatic amino acids as in some meteorites. The strongest bands of each amino acid are also found in the combined spectrum of >30 interstellar locations in diverse star-forming regions supporting the suggestion that amino acids are widely spread in interstellar space. Future mid-IR searches for proteinogenic amino acids in protostars, protoplanetary disks and in the interstellar medium will be key to establish an exogenous origin of meteoritic amino acids and to understand how the prebiotic conditions for life were set in the early Earth.**

## 1. INTRODUCTION

Proteinogenic and non-proteinogenic amino acids have long been known to be present in meteorites, especially carbonaceous chondrites (Kvenholden et al. 1970, Cronin and Moore 1971, Cronin & Chang 1993, Ehrenfreud & Charnley 2000) and their abundances have been compiled in numerous studies (e.g. Commeyras et al. 2005, Martins & Sephton 2009, Martins 2011, Cobb & Pudritz 2014, Pizzarello & Shock 2017; Koga & Naraoka 2017). The presence of amino acids in meteorites and comets (Elsila et al. 2009) suggests that the young planet Earth was enriched with these molecules, basic constituents of proteins, essential for the emergence of life and the development of living organisms (Chyba et al 1990, Botha and Bada 2002, Muñoz Caro et al. 2002).

A few dozen molecules have been detected in the interstellar medium of star-forming regions (Cazaux et al. 2003, Jorgensen et al. 2012, Kahane et al. 2013, Favrè et al. 2018) and in protoplanetary disks (Dutrey et al. 2014). These detections include aldehydes, acids, ketones, and sugars. The simplest organic acid, formic acid (HCOOH), which contains the carboxyl group, one of the main functional groups of amino acids, has been detected in low-mass star-forming regions (Liu et al. 2002, Lefloch et al. 2017) and in a protoplanetary disk (Favrè et al. 2018). The reaction of protonated alcohols and HCOOH could lead to the production of glycine (and other amino acids) in the hot nuclei of molecular clouds (Ehrenfreud and Charnley 2001) and formation of amino acids in the interstellar medium is posible via specific gas-phase reactions in dark clouds (Ehrenfreund and Charnley 2000, Elsila et al. 2007, Martins and Sephton, 2009, Redondo et al. 2017). This suggests that amino acids could be present in star-forming regions, however, conclusive evidence has not been obtained yet. Very sensitive searches of amino acid bands conducted at millimeter wavelengths have focused on glycine, the most simple amino acid, and obtained upper limits to the column density (e.g. Ceccarelli et al. 2000) or tentative identifications of ro-vibrational bands (Kuan et al. 2003) which have not been finally confirmed (Snyder et al. 2005, Jones et al. 2006, Cunningham et al. 2007). Identifying the site and processes of amino acid formation in interstellar space remains a challenge of great importance in astrochemistry and astrobiology.

The mid-IR spectrum of amino acids offers an alternative way to explore their presence in the interstellar medium. Systematic laboratory work (e.g. Husain et al. 1984, Matei et al. 2005, Iglesias-Groth and Cataldo 2018) has led to the identification of a large number of relatively intense amino acid bands in the mid-IR. Between 15 and 20.8 μm these bands are mostly due to the carboxylate group $COO^-$ rocking, bending and wagging modes; in the range 20.8 to 26.3 μm are mainly associated to the protonated amino group, i.e. $NH_3^+$ vibration modes and, at longer wavelengths, between 26.3 and 37.0 μm the bands are mainly related to the $C-C^{\alpha}-N$ deformation modes (see e.g. Iglesias-Groth and Cataldo 2018). Weak hydrogen bond vibration bands have also been identified from 45.5 to 100.0 μm which may be useful to achieve a solid amino acid identification. The *Spitzer Space Telescope* has obtained a large, very valuable number of moderately high resolution spectra of star-forming regions in the spectral range 10-40 μm. This spectral range appears particularly interesting for a potential identification of individual amino acid species since band patterns significantly vary among amino acids (see e.g. Iglesias-Groth and Cataldo 2018).

The young stellar cluster IC 348 (age ≈ 2 Myr), located at the eastern end of the well known Perseus Molecular Cloud complex (Herbig 1998, Luhman et al. 2003, Bally et al 2008, Luhman et al. 2016) at a distance of 321 ±10 pc (Ortiz-León et al. 2018), is one of the nearest star-forming regions which has proved well suited to explore the presence of complex carbon-based molecules (see. e.g. Iglesias-Groth 2019). The Perseus Molecular Cloud also provided one of the first unambiguous detections of the anomalous microwave emission (Watson et al. 2005), a radiation likely caused by electric dipole emission (Draine and Lazarian 1998) of a diffuse distribution of fast spinning carbonaceous molecules (polycyclic aromatic hydrocarbons, hydrogenated fullerenes, etc.). Organic molecules appear to be widely distributed in this nearby molecular cloud complex which prompted us to carry out an extensive search for amino acids.

Among all the 20 proteinogenic amino acids precise laboratory measurements of wavelengths, molar extinction coefficients and integrated molar absorptivities of mid-IR bands are available for the aromatic amino acids tryptophan, tyrosine and phenylalanine and for the aliphatic isoleucine and glycine (Iglesias-Groth and Cataldo 2021). In this work, we present results on the search in the Perseus Molecular Complex for mid-IR bands (10-40 μm range) of these five amino acids.

## 2. OBSERVATIONS

We have used high spectral resolution (R≈600) archive spectra obtained with the

Infrared Spectrograph (IRS) onboard the *Spitzer Space Telescope* at various interstellar medium locations in the central region of IC 348 in the Perseus Molecular Cloud. All the selected observations were located within 10 arcmin of the most luminous star of the cluster, HD 281159, within the region marked in Fig. 1. Fully reduced spectra acquired with the Short-High (S-H, 9.8-19.5 μm ) and Long-High (L-H, 19.5-36 μm) modules were taken from the Cornell Atlas of Spitzer/IRS Sources (CASSIS; http://cassis.sirtf.com, Lebouteiller et al. 2015) which provides reduced and flux calibrated data. The spectra used from both modules of the IRS are identified with the corresponding AOR numbers in Table 1. Coordinates of the ISM pointings are also listed. The present search is focused on interstellar locations so we have selected pointings which in the original observing programmes, mainly focused on protoplanetry discs, where serving as background observations for nearby targets. Our interest in this first work is to explore the presence of amino acids in the ISM as disc continuum emission could limit the detectability of faint bands.

The S-H and L-H spectra are not available for the same interstellar locations, however we do not intend to carry out studies of individual locations, on the contrary, we will average the available spectra from different pointings in the core region of IC 348 from both IRS modules in order to gain as high signal to noise as possible to facilitate the search for weak bands. For comparison purposes, 32 spectra obtained with the same instrument at ISM locations in various star-forming regions, unrelated to the Perseus molecular complex, will be combined to generate a star-forming region "ISM reference" spectrum of similar high signal to noise to that of Perseus. The log of these observations can be found in Table 2 which lists AOR numbers and pointing coordinates for these Spitzer spectra. In this case, for each pointing there is full spectral coverage from 10 to 35 μm.

## 3. DATA

The CASSIS pipeline takes the Basic Calibrated Data (BCD) images, the associated uncertainty images and the bad pixel mask as starting products. The BCD images are produced by the Spitzer Science Center BCD pipeline. The High-resolution modules of the IRS are significantly affected by cosmic ray hits and spurious features, thus the CASSIS pipeline (Lebouteiller et al. 2015) pays special care to the exposure combination and image cleaning to remove any bad pixels in order to minimize these effects in high resolution data. This pipeline incorporates various improvements with respect to Lebouteiller et al. (2011) including several techniques for optimal extraction and a differential method that eliminates low-level rogue pixels.

As we are interested on observations of the general interstellar medium we will use the CASSIS full-aperture extraction for extended sources. In order to compute fluxes, this extraction method co-adds the pixels in the detector area corresponding to one wavelength value. Since the flux is integrated, the presence of bad pixels anywhere within this area is especially damaging and bad pixels need to be picked out and replaced. The "cleaning" is carried out on the combined image of all exposures obtained for a given target since bad pixel fluxes are substituted using neighbours whose flux is more reliable when exposures have been combined. The most problematic are the rogue pixels which have elevated dark current that change unpredictably with time. The CASSIS pipeline uses IRSCLEAN, (http://irsa.ipac.caltech.edu/data/SPITZER/docs/dataanalysistools/tools/

irsclean) an IDL tool for creating bad pixel masks from Spitzer IRS BCD (and pre-BCD) image data, and "cleaning" the masked pixels in a set of data to substitute bad pixels.  As described by Leboutellier et al. (2015) this includes pixels with no values (NaNs), with a high bad pixel mask value (>256), bad pixels and rogue pixels flagged in the campaign mask, pixels with a large uncertainty and negative pixels.

The new pixel values calculated by the IRSCLEAN algorithm are mainly based on neighbouring pixels taking into account uncertainties and mask values propagated for each step. Subsequently, the full-aperture extraction is performed using the standard tool in SMART. An error on the flux is ultimately calculated using the quadratic sum of the pixel uncertainties in the detector area corresponding to each wavelength value. No background subtraction is applied for full-aperture extraction, however this is not expected to introduce any significant uncertainty on the measurement of wavelengths and fluxes of the relatively weak bands which we aim to detect.

As a check of the effectiveness of the bad pixel correction used in the full aperture extraction technique, we have compared the CASSIS full aperture and optimal differential extracted spectra for the well studied source RNO 90 ( Pontoppidan et al 2010 , Salyk et al. 2011).  Figure 2 shows the spectra obtained by these two CASSIS extraction methods, small flux differences (typically less than 0.05 Jy) are seen in the continuum, probably due to the background correction applied by the optimal differential extraction. The corrections for bad pixels that both extraction techniques use do not seem to produce residual artifacts which be confused with weak bands. If this happens, such artificial bands would peak-fluxes significanty below 0.05 Jy.  As a further test we have measured fluxes for well known water bands in the full aperture spectrum of RNO 90 and compared them (see Table 3) with those reported by Blevins et al. (2016) from a completely indepedent reduction of the Spitzer RNO 90 spectrum.  The comparison of band fluxes shows good agreement at the level of 1 W m$^{-2}$ .

The individual CASSIS full-aperture extraction of the S-H and L-H spectra available in the core of IC 348 showed very similar features and were subsequently averaged to produce a high signal to noise spectrum designated hereafter as "combined IC 348 ISM". Similarly, we produced a star-forming region ISM "reference" spectrum by averaging the 32 spectra selected from ISM observations of other diverse star-forming regions. The minimum flux level for a 3σ detection of a line in the combined IC 348 ISM spectrum is determined at 0.5 x 10$^{18}$ W m$^{-2}$. Line detectability at such low fluxes opens the possibility to identify weak transitions, enabling, in particular, a search for mid-IR transitions of the selected amino acids.

## 4. RESULTS

Laboratory wavenumbers, wavelengths (laboratory and observations) and integrated molar absorptivities (ψ) for the strongest mid-IR transitions of the amino acids under study are listed in Table 4. Laboratory wavenumbers are precise at the level of 1 cm$^{-1}$ and integrated molar absorptivities (adopted from Iglesias-Groth and Cataldo 2021) have uncertainties of order 20%. In Figures 3-8 the positions of the strongest laboratory bands of each of the five amino acids under study and the emission features we tentatively assigne to them are marked on the combined IC 348 ISM spectrum (red-line). For comparison, in each figure it is also plotted the star-forming region ISM "refeence" spectrum (black line). For all the strongest mid-IR bands of amino acids we have identified potential emission features in both spectra with wavelenghts fully consistent within errors with those measured at laboratory.

Both, the IC 348 and the reference ISM spectra show a very large number of emission lines in common. The most intense lines are due to well identified molecular and atomic transitions. The strongest and broader feature in the Perseus spectrum is the 11.2 μm PAH band (see e.g. Fig. 3, top panel). Other strong lines which we can see in the panels of Figures 3, 4 and 5 are associated to transitions of molecular hydrogen at 12.28 and 17.04 μm, and atomic hydrogen at 12.37 and 19.06 μm (marked in the figures). Weaker lines of water (marked with blue asterisks), fullerene species and other organic molecules are also seen in the spectra.

Fluxes for the assigned bands were measured only in the IC 348 ISM spectrum by integrating the continuum subtracted spectrum. In Table 4 we list, for each of the five amino acids, measured fluxes and wavelengths for all the laboratory bands with integrated molar absorptivities $\psi > 2$ km mol$^{-1}$. Errors in the band fluxes were estimated by propagating pixel error through the band integrals and are typically of order 20%. We tentatively identify 14 lines for tryptophan, 12 lines for tyrosine, 7 lines for phenylalanine and 5 lines for isoleucine and glycine, respectively.

Tryptophan. In Table 4 there are 12 bands with reported integrated molar absorptivities $\psi > 2$ km mol$^{-1}$, we identify in Fig. 3 emission features which could be associated to each of these laboratory bands and provide fluxes for each of them. We also list fluxes for emission lines that coincide with other reported bands of tryptophan for which there are no available molar absorptivities. The two bands with larger integrated molar absorptivity ($\psi > 30$ km mol$^{-1}$) have clear counterparts at 13.45 and 19.64 μm. There are also four laboratory bands with $\psi > 10$ km mol$^{-1}$ which coincide with emission features in the spectra at 11.58, 17.88, 23.47 and 28.70 μm. Other bands with low values of $\psi = 2-5$ km mol$^{-1}$ seem to be present in the spectrum albeit at more modest fluxes. For any reported laboratory band with a measured $\psi$ we find a counterpart emission feature in the observed spectrum as marked in the figure.

Tyrosine. All the 12 bands listed in Table 4 have a counterpart in the observed spectrum (see Fig. 4). There are six bands with $\psi > 10$ km mol$^{-1}$ but two of them at 17.36 and 18.86 μm are not plotted because they are severely blended with well known water and fullerene bands. The band at 26.32 μm is marginally resolved (the observed feature is too broad) and very likely contaminated by an unknown specie at a slightly longer wavelength. The other six remaining weak bands of tyrosine have their proposed counterparts marked in Fig. 4 and fluxes reported in the table.

Phenylalanine. All eight lines listed in Table 4 have tentative counterparts marked in Fig. 5. The bands with the two higher $\psi$ values ($\psi > 20$ km mol$^{-1}$) at 14.29 and 27.31 μm also have the higher measured fluxes. The line at 27.31 ($\psi > 50$ km mol$^{-1}$) appears however with a red tail which indicates some contamination. The line at 14.29 μm is weaker but with a well defined profile. The other bands have lower ψ values and also the assigned lines in the spectrum present lower fluxes, but overall the evidence for the presence of this amino acid appears sound.

Isoleucine. The eight bands listed in Table 4 have low ψ values with the highest, ψ = 13 km mol$^{-1}$ , corresponding to the 18.58 μm band which also shows the highest flux observed flux in the proposed counterpart emission features. Three of the other seven bands are clearly seen in Fig. 6 but are unfortunately blended with other molecules and two of the remaining bands have low fluxes at only 2 sigma from the detectability limit making. The evidence for this amino acid is less solid than the previous ones.

Glycine. There are only six known bands in the spectral range under consideration, out of which five have integrated molar absorptivities reported. The two bands with higher ψ values at 19.85 and 27.75 μm and the 14.32 μm band can be assigned to rather isolated emission lines in the IC 348 spectrum. The band at 10.2 μm is blended with the strong PAH band and its presence (and flux) is difficult to ascertain. The remaining band at 16.45 μm could be associated with an emission line which is however superimposed on a weak broader spectral feature and the measured flux is rather uncertain. The evidence for the presence of glycine in the ISM of IC 348 relies on the proper identification of only a few availabe bands andshall be taken with caution.

The bands with the largest laboratory integrated absorptivities generally stronger emisión lines counterpart in the observed spectra, both in Perseus and in the reference spectrum. Emission features likely contaminated by water lines are indicated in Table 4, but we caution that emission features assigned to amino acids may also be contaminated by other unknown species. Future high resolution mid-IR observations should clarify the extent of contamination . The vibrational excitation diagrams (see below) suggests that if this contamination exists in some bands, it is not a dominant contribution to the flux, otherwise the correlations would be lost.

## 5. DISCUSSION

*5.1 Exploring excitation diagrams and equilibrium temperatures*

In order to produce vibrational excitation diagrams it is necessary to estimate for each amino acid transition the number of molecules in the upper vibrational state $N_u$. This can be obtained from the measured flux of each transition assuming optically thin emission, since the total power emitted in a band (see e.g. Camí et al. 2010) is $P=N_u A_{ul} h \nu_{ul} / 4\pi D^2$ , where, D is the distance to IC 348 (321 pc), $\nu_{ul}$ the frequency of the amino acid transition, and h the Planck constant. The Einstein $A_{ul}$ coefficients can be obtained from the laboratory integrated molar absorptivities, which ideally should be measured in conditions ressembling those of the astrophysical environment under study. As molar absorptivity coefficients are sensitive to temperature and to vaccuum conditions, we can only expect from the avalaible absortivity data a very preliminary indication of the existence of termal equilibrium.

In spite of the limited laboratory information available, vibration excitation diagrams for each amino acid were obtained using the strongest transitions for which molar absorptivities were available and fluxes measured in the "combined IC 348 ISM" spectrum (listed in Table 4). The results, plotted in Figure 8, show good correlation coefficients in the range r=0.9-0.95 for each individual amino acid, even if for simplicity a vibrational degeneracy $g_u$=1 was adopted for all the energy levels. The inverse of the slopes of the fits in the figure indicates equilibrium temperatures in the range 270-290 K for each amino acid. Fortunately, this is very close to the temperature of the laboratory setting which provided the integrated molar absorptivities. This suggests that amino acids can exist in rather warm gas of the stellar cluster. Values for total number of emitting molecules of amino acids are found in the range $N_{tot}= 10^{42}$ from the ordinate at origin, this shall be taken only as order of magnitude indicative given the various approximations and limitations involved. For comparison, fullerene C60 and C70 estimates in IC 348 based on Spitzer spectra obtained with the same instrument at similar locations in the core of the IC 348 cluster led to equilibrium temperature estimates of order 200 K and $N_{tot}$ values about one hundred times higher (Iglesias-Groth 2019).

*5.2 Column densitities*

Assuming that amino acids are present in the ISM of IC 348 we may attempt to estimate their column densities. Considering that the absorbed UV energy is re-emitted mostly via the IR vibrational bands, column densities, n(AA), can be derived from the measured band fluxes in the IC 348 spectrum. Reliable absolute abundances would require molar absorptivities of the mid-IR bands measured in conditions properly ressembling the ISM in Perseus, the available laboratory data (obtained at approx. 300 K) may not be ideal, but can provide some useful insight on the relative abundances of the amino acids under consideration. We will follow procedures similar to those used to estimate abundances for other molecular species in the ISM of star-forming regions (e.g. Berné et al. 2017, Iglesias-Groth 2019). The total IR intensity (W $m^{-2}$ $sr^{-1}$) emitted by the amino acid AA can be estimated as $I_{tot}$= n(AA) x $\sigma_{UV}$ x $G_0$ x 1.2 x $10^{-7}$ where $G_0$ is the radiation field in the locations of the observations and $\sigma_{UV}$ is the cross section for absorption in the UV of the relevant molecular specie.

The standard value of the radiation field in the general ISM (Habing 1968), $G_0$=1, corresponds to 1.2 W $m^{-2}$ $sr^{-1}$. Large variations exist in the far UV radiation field within the IC 348 star-forming region (Habing 1968). As the location of the observations under consideration are relatively close to the positions of the most luminous stars of the cluster, the appropriate value of $G_0$ wil be significatly higher. Previous work on fullerene abundances using spectra obtained in very similar locations of the ISM in IC 348, adopted $G_0$ = 45 for the average interstellar radiation field ( Iglesias-Groth 2019). The same value will be adopted here.

The total intensity, $I_{tot}$, can be determined from the measured total fluxes emitted in the IR. However, we only measured fluxes for the subset of stronger lines known in the 10-30 μm region. The real total IR intensity will be higher than the value inferred from our band measurements, and subsequently, the column densities too. The excitation diagram suggests that amino acids in the ISM of IC 348 are in thermal equilibrium with exctitation temperatures in the range 270-290 K. We will assume such thermal equilibrium conditions to model the contribution to the total IR flux resulting from the known non-measured IR transistions (out of the range 10-30 μm) for which molar absorptivities are available (Iglesias-Groth and Cataldo 2021) and apply an upward correction to the total flux for each molecular specie. The resulting correction factors are 2 for tryptophan, 3 for tyrosine and isoleucine and 4 for glycine and phenylalanine.

The estimated total emited IR fluxes after applying these corrections are 10.6, 8.6 and 0.8 x $10^{-17}$ W m$^{-2}$ for the aromatic tyrosine, phenylalanine and tryptophan respectively, and 5.2 and 12.9 x $10^{-17}$ W m$^{-2}$ for the aliphatic isoleucine and glycine, respectively. As we have used full aperture extraction spectra, band fluxes are converted into I$_{tot}$ dividing by the subtended area in the sky by the corresponding slit of each module of the IRS spectrograph.

The third parameter needed to infer column densities is the amino acid UV absorption cross section, σ$_{UV}$. This can be derived from molar absorptivities. The aromatic amino acids contain conjugated aromatic rings and therefore are very efficient absorbing light in the UV range. In fact, absorption in the range 200-300 nm is dominated by the aromatic side-chains of tryptophane, tyrosine and phenylalanine. UV absorption cross sections σ(cm$^2$) can be computed from laboratory measurements of molar absorption coefficients ε (λ) using the relation σ(λ)= 1000 ε (λ)/(N$_A$ log (e)) where N$_A$ is the Avogadro number, and log (e) is the decimal logarithm of the Euler number. In Fig. X we display molar absorption coefficients ε (mol$^{-1}$ cm$^{-1}$) for the five amino acids under consideration. The plot shows molar absorption coefficient values as a function of wavelength for the range of interest 200-300 nm.

In the IC 348 star-forming region the most luminous stars dominating the UV spectrum are of type B5 and A2 with effective temperature (of 15000 and 10000 K, respectively, and the peak of their combined emitted radiation is around 200 nm. At shorter wavelengths the flux emitted by these stars sharply decreases. More than 400 cooler low-mass stars have been identified (Luhman et al. 2016 ) in IC 348, but their luminosities are much lower and the peak of their emitted radiation occurs at wavelenghts longer than 300 nm where the amino acids under study have negligible absorbances (see Figure 9). Thus, given the expected stellar radiation field the most relevant wavelength range for determining effective UV absorption cross sections is from 190 to 230 nm. This range is determined by the radiation emitted by the most luminous stars in IC 348. The resulting effective molar extinction coefficients, **ε$_{eff}$** are 10850, 5640, 3530, 90 and 50 mol$^{-1}$ cm$^{-1}$, for tryptophan, tyrosine, phenilalanyne, isoleucine and glycine, respectively. Note the very low molar extinction coefficients of glycine and isoleucine as compared with those of the aromatic amino acids. Mean UV absorption cross-sections are estimated in this work using these **ε$_{eff}$** values and the expression above . Adopting these values as reference (Fasman 1976) we obtain for the corresponding UV absorption cross sections 4, 2, 1, 0.03 and 0.02 x $10^{-17}$ cm$^2$, for tryptophan, tyrosine, phenylalanine, isoleucine and glycine, respectively.

The resulting column densities for the aromatic amino acids are then: phenylalanine n(Phe)= 1 x $10^{11}$ cm$^{-2}$, tyrosine n(Tyr)= 0.8 x $10^{11}$ cm$^{-2}$ , and tryptophan n(Trp)= 0.6 x $10^{11}$ cm$^{-2}$. For isoleucine we find a much higher column density n(Ile)= 2 x $10^{12}$ cm$^{-2}$ . For glycine, the corresponding column density based on the few transitions available in our spectral range appears to be the highest n(Gly) = 9 x $10^{12}$ cm$^{-2}$. We recall that estimates of the latter two amino acids plausibl relay on the identification of only a few bands. Additional observations of glycine and isoleucine in Perseus using transitions in other spectral ranges would be highly valuable. An upper limit to the glycine column density of 5 x $10^{12}$ cm$^{-2}$ was established for the low-mass protostar IRAS 16293-2422 (Ceccarelli et al. 2000) in a different star-forming region, it would be important to carry out a similar sensitive search for millimetric transitions of this amino acid across the Perseus Molecular Complex.

*5.3 Relative abundances*

Glycine is the most abundant amino acid in any type of carbonaceous chondrites with abundances of order 5-6 ppm in CM2 types and 200 ppm in CR2 types, the meteorites with a higher abundance of amino acids. In both types, isoleucine, phenylalanine and tyrosine are also systematically detected although at lower concentrations (Coob & Pudritz 2014), with abundance levels of order 1, 0.8 and 0.9 ppm, respectively, for CM2 types. In CR2 types, isoleucine is more abundant with values of order 30 ppm, followed by phenylalanine 20 ppm, while tyrosine is at the level 0.7 ppm. Among the amino acids studied in this work, glycine and isoleucine are the most abundant in CR2 meteorites. They appear to be also the most abundant in the ISM of the IC 348 star-forming region. On the other hand tryptophan, which we find to be the less abundant of the five studid amino acids in Perseus, interestingly it is undetected in meteorites.

Another relevant comparison with meteorites is the fraction of gas phase locked carbon in amino acids. This can be estimated from the amino acid column densities via $f_C(AA) = n(AA) \times N_{aa} / N(H) \times [C]$, where $[C]=1.6 \times 10^{-4}$ is the carbon to hydrogen ratio in the región ( Sofia et al. 2004), $N_{aa}$ is the number of carbon atoms in the amino acid and, N(H) is the hydrogen column density, $4.8 \times 10^{22}$ cm$^{-2}$ for the central part (Sun et al. 2008) of IC 348. The results are $f_C$ = 0.08, 0.09, 0.1, 1.8 and $2.4 \times 10^{-6}$ for tryptophan, tyrosine, phenylalanine, isoleucine and glycine, respectively. It is remarkable that these values for isoleucine and glycine are very close to the 3 parts per million versus carbon reported in several types of carbonaceous chondrites (Cronin & Chang 1993, Botha & Bada 2002).

The emission features assigned to amino acids in the ISM spectrum of Perseus are also present in the combined ISM spectrum from star-forming regions. Column densities cannot be inferred from this spectrum because the sources are distributed in a large range of distances and fluxes would not be meaningful, however the spectrum clearly displays the same bands assigned to amino acids with relative intensities similar to those found in IC 348. Suggesting that amino acids are widely distributed across the Galaxy.

**6. CONCLUSIONS**

We have conducted a search for mid-IR (10-30 μm) transitions of five amino acids (tryptophan, tyrosine, phenylalanine, glycine and isoleucine,) in the Spitzer spectrum obtained averaging observations from various ISM locations in the core of the IC 348 star forming cluster in the Perseus Molecular Cloud. For each of the strongest laboratory bands of these amino acids have a counterpart emission feature in this spectrum. Under the assumption that these emission features bands are mainly produced by IC 348 ISM amino acids being at it is proposed have built preliminary vibrational excitation diagrams that are consistent with equilibrium temperatures in the range 270-290 K. This should be taken cautiously as contamination of the emission features from other unknown molecular species it is posible and shall be further investigated. From the available UV absorption cross sections of these amino acids and the total estimated IR fluxes in amino acid bands, preliminary column densities have been obtained. Interestingly, we find the relative abundance pattern of these molecules in the ISM of IC 348 to be similar to that known in meteorites, with glycine and isoleucine being much more abundant than any of the other three aromatic amino acids.

If, as suggested by these findings, amino acids are present in the ISM of star-forming regions, they could also be part of the inventory of organic molecules in protoplanetary disks. Searches in protostars and protoplanetary disks in Perseus and other molecular cloud complexes are worthwhile as they should provide valuable insight on the delivery of complex organics by meteoritic and cometary material to planets in early stages of formation and, ultimately, on the processes relevant to the origen of life on Earth.

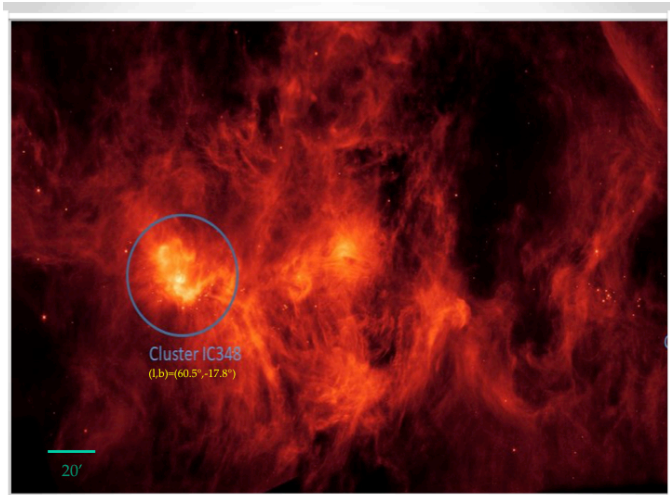

**Fig. 1**
The Perseus Molecular Cloud as observed by NASA´s Spitzer Space Telescope showing the location of the IC 348 star cluster and encircled the region containing the IC 348 pointings listed in Table 1. NASA/JPL-Caltech - https://photojournal.jpl.nasa.gov/figures/PIA23405_fig2.jpg

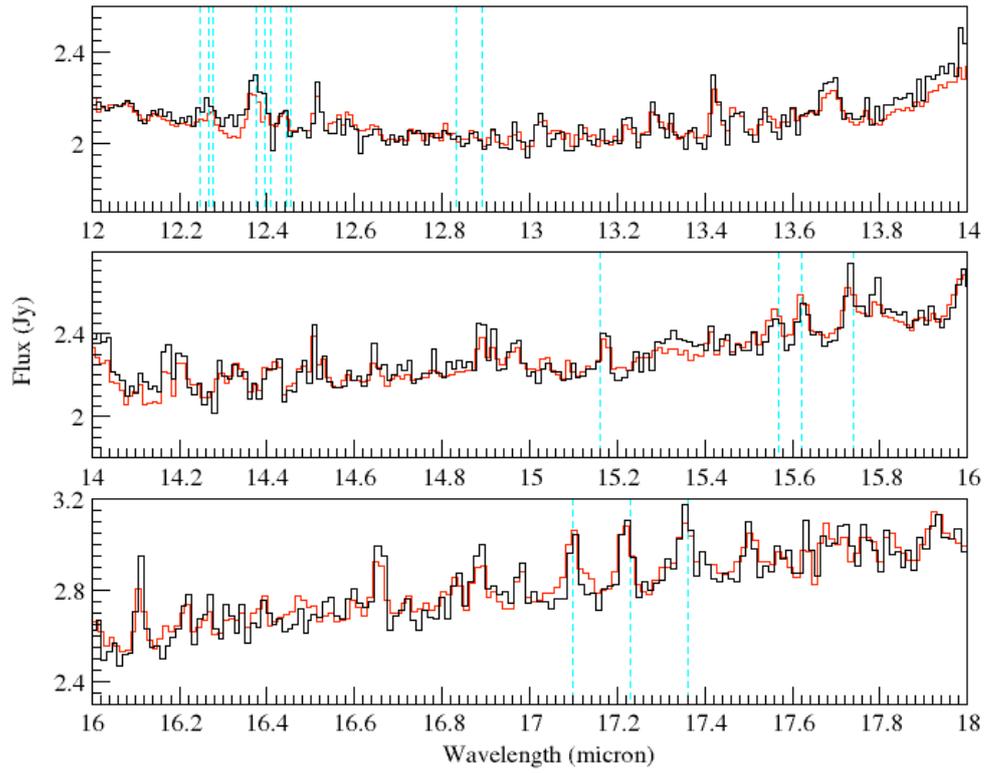

Figure 2  CASSIS spectra of RNO 90  provided by the full aperture  and optimal differential  extraction techniques.

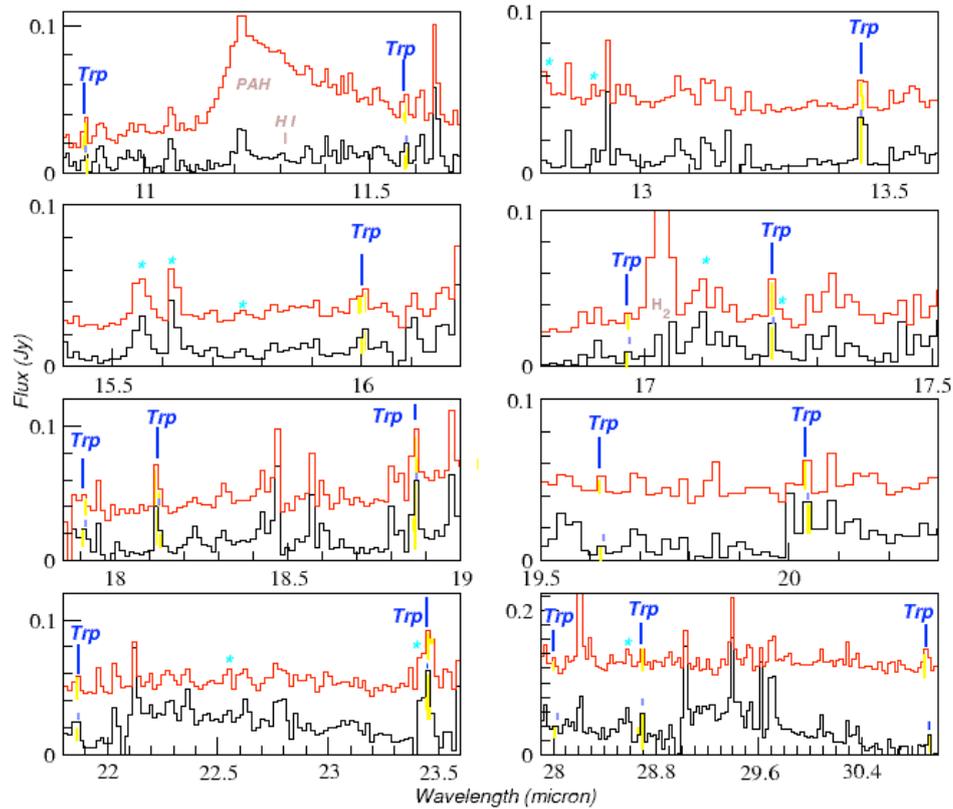

Figure 3. Tryptophan (Trp) bands: laboratory wavelengths are marked with vertical lines. Emission features assigned to these bands are indicated (yellow filled) in the "combined IC 348 ISM" spectrum (red colour) and in the "combined star-forming region ISM" spectrum (black). Blue asterisks mark the position of water bands. The spectra are shifted on the vertical axis for convenience of display, only band fluxes relative to the local continuum can be deduced from these plots.

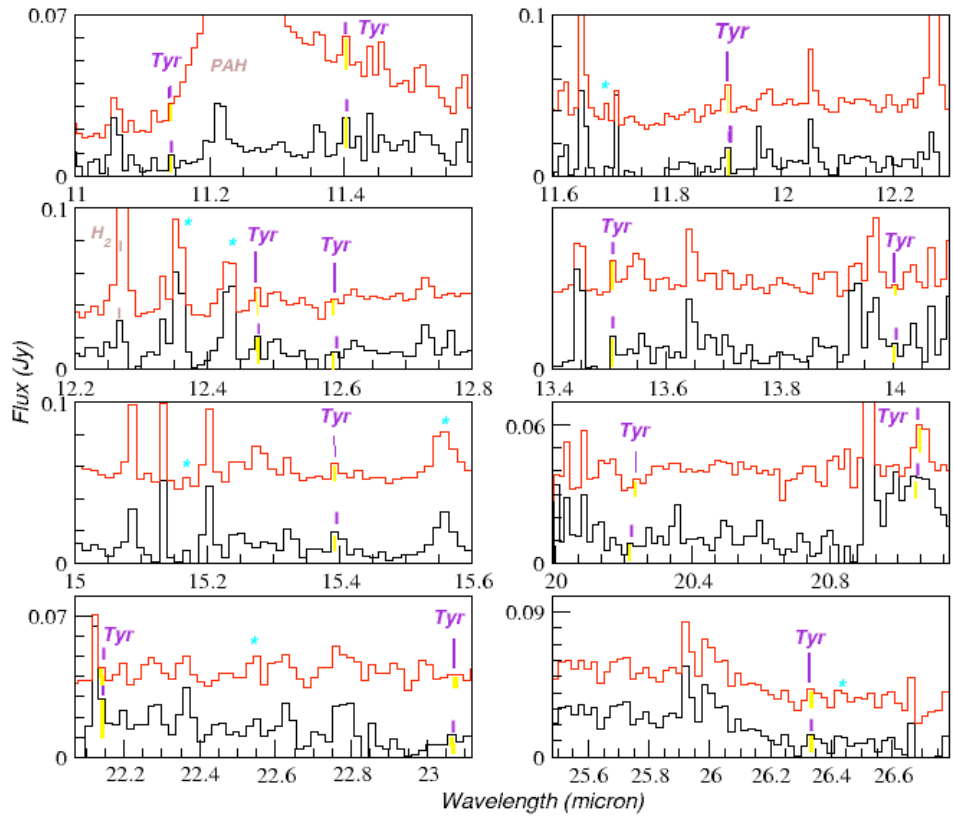

Figure 4. Tyrosine (Tyr) bands: caption as in Fig. 3.

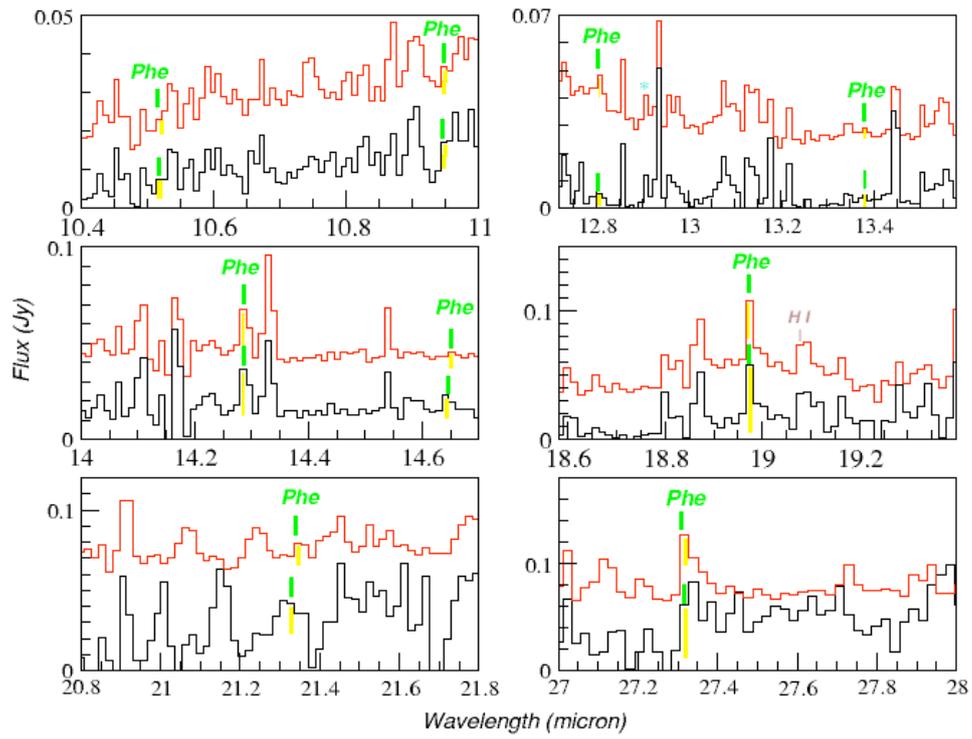

Figure 5. Phenylalanine (Phe) bands: caption as in Fig. 3.

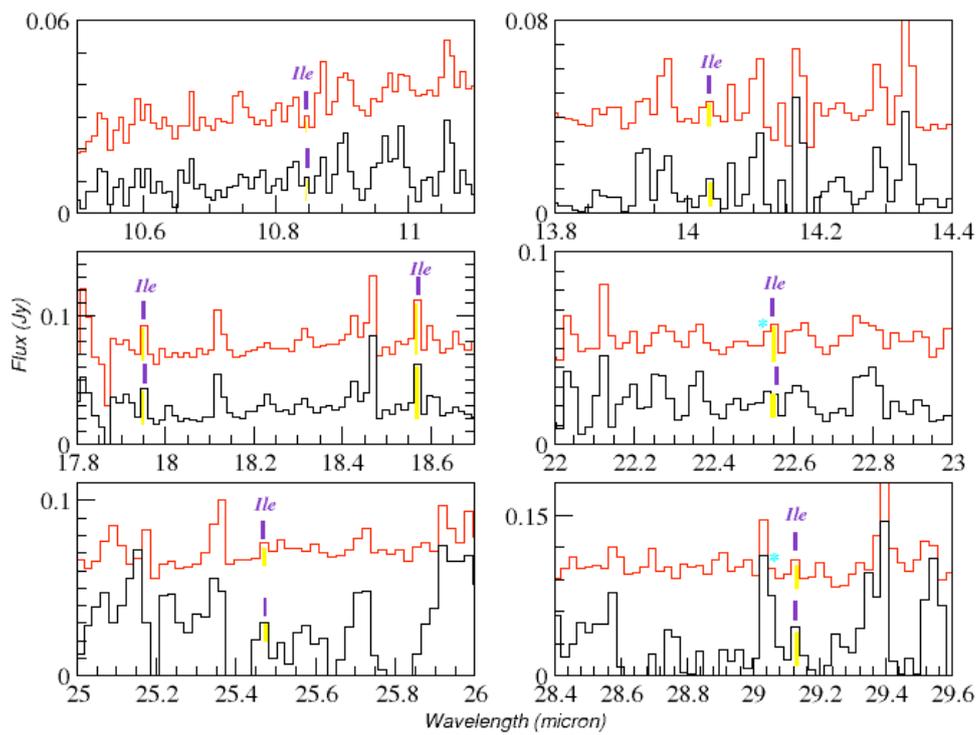

Figure 6. Isoleucine (Ile) bands: caption as in Fig. 3.

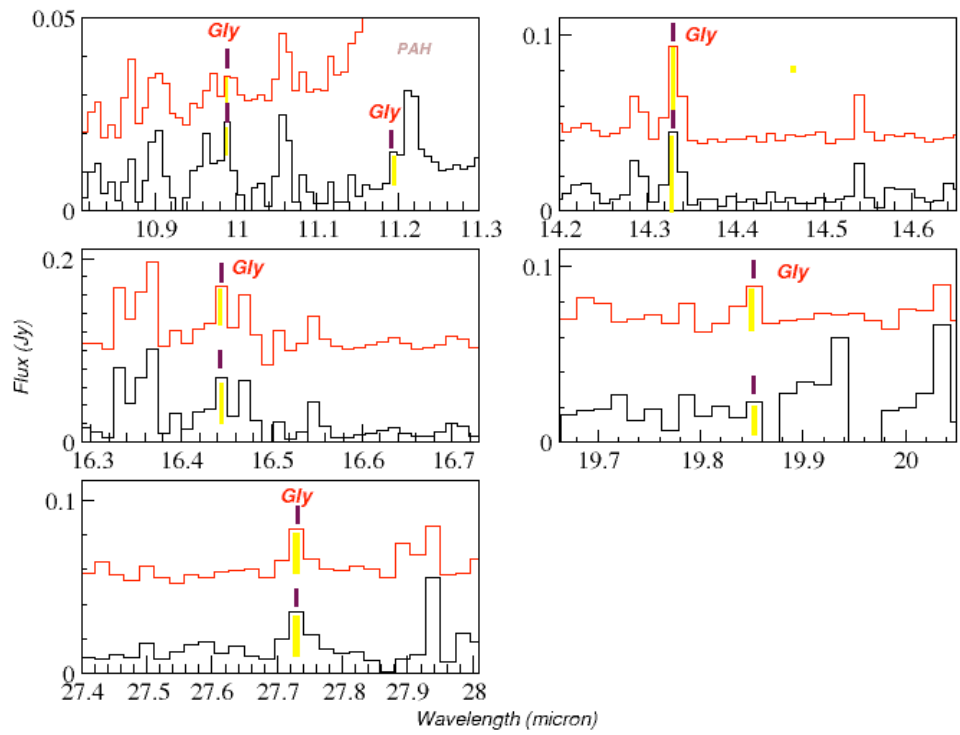

Figure 7　Glycine (Gly) bands: caption as in Fig.3

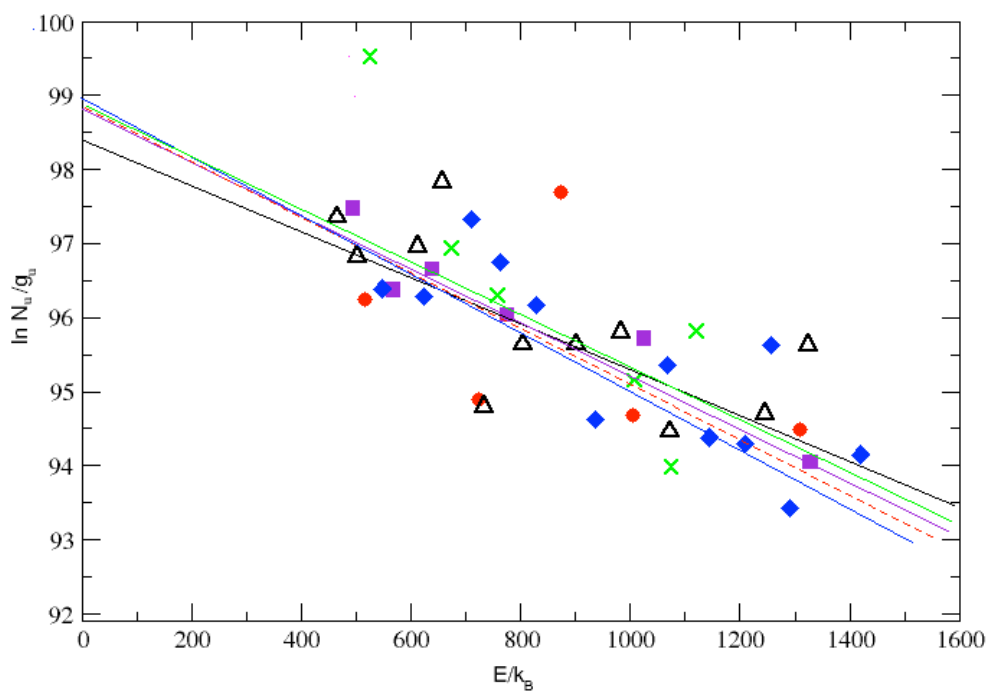

**Figure 8**. **Vibrational excitation diagram for the bands of three aromatic amino acids, isoleucine and glycine observed in the ISM of IC 348 (Perseus Molecular Cloud).** Natural logarithm of $N_u/g_u$ versus energy $E_u/k$ of the excited vibrational states for all the observed amino acid bands with available fluxes and integrated molar absorptivities. Tryptophan transitions (black triangles), tyrosine (blue diamond), phenylalanine (green crosses), isoleucine (violet squares) and glycine (red circles).

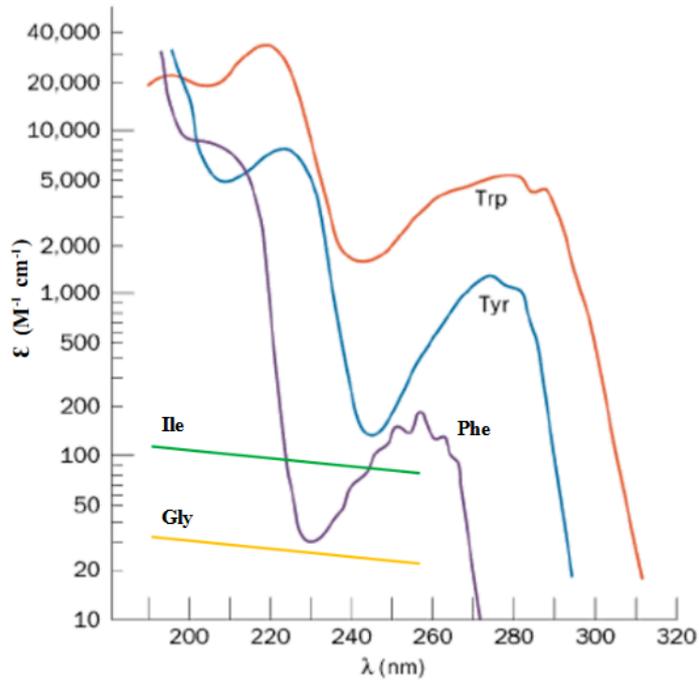

**Figure 9.** UV absorbance of amino acids. Molar extinction coefficients as a function of wavelength (Wetlaufer 1963).

**Table 1. IC 348 ISM locations observed with Spitzer IRS**

Short-High spectra

| AOR | R.A.$_{2000}$ | Dec$_{2000}$ |
|---|---|---|
| 22848000[c] | 03:44:38.90 | +32:10:03.0 |
| 22848512[c] | 03:44:41.66 | +32:06:45.4 |
| 22849024[c] | 03:45:00.87 | +32:09:15.5 |
| 22849536[c] | 03:44:22.88 | +32:04:57.0 |
| 22850048[c] | 03:44:42.71 | +32:03:29.8 |
| 22850560[c] | 03:44:36.08 | +32:00:14.7 |
| 22851072[c] | 03:43:50.10 | +32:08:17.9 |
| 22851584[c] | 03:44:33.23 | +31:59:54.1 |

Long-High spectra

| AOR | R.A.$_{2000}$ | Dec$_{2000}$ |
|---|---|---|
| 27542016_4[w] | 03:44:32.67 | +32:11:05.2 |
| 27542016_2[w] | 03:44:30.31 | +32:11:35.2 |
| 27542016_3[w] | 03:44:27.95 | +32:12:05.2 |
| 27542016_1[w] | 03:44:30.31 | +32:11:35.2 |

[c] program ID 40247 (N. Calvet, P.I.)
[w] program ID 50560 (D. Watson, P.I)

**Table 2. Spitzer spectroscopic observations of various ISM locations**

| AOR | R.A.$_{2000}$ | Dec.$_{2000}$ |
|---|---|---|
| 27058432 [c] | 04:33:43.51 | +25:20:38.1 |
| 24404480 [n] | 04:35:27.37 | +24:15:58.9 |
| 27062784 [c] | 16:26:23.42 | -24:20:00.5 |
| 27057152 [c] | 04:42:41.01 | +25:15:37.3 |
| 27067136 [c] | 04:47:11.45 | +16:58:42.9 |
| 27059712 [c] | 18:30:04.90 | +01:15:06.5 |
| 27063808 [c] | 16:02:57.80 | -40:18:25.3 |
| 27062528 [c] | 04:14:14.59 | +28:28:43.0 |
| 27060224 [c] | 04:23:42.12 | +24:56:14.3 |
| 27058688 [c] | 04:23:39.18 | +24:56:14.3 |
| 27060224 [c] | 04:23:42.12 | +24:56:14.3 |
| 27062016 [c] | 16:31:33.46 | -24:28:37.4 |
| 27059200 [c] | 04:31:53.51 | +24:24:17.8 |
| 27059200 [c] | 04:31:53.51 | +24:24:17.8 |
| 27057664 [c] | 04:35:50.22 | +22:49:41.9 |
| 27064832 [c] | 15:45:08.02 | -34:18:15.8 |
| 27064320 [c] | 15:56:03.25 | -37:56:06.3 |
| 27063552 [c] | 03:30:44.01 | +30:33:47.0 |
| 27061760 [c] | 16:34:05.01 | -15:48:16.8 |
| 27064064 [c] | 15:56:42.31 | -37:49:15.4 |
| 27066112 [c] | 11:08:15.49 | -77:41:43.6 |
| 27065856 [c] | 11:09:36.51 | -76:23:20.7 |
| 27066112 [c] | 11:08:15.49 | -77:41:43.6 |
| 24402944 [n] | 11:01:55.52 | -34:43:02.1 |
| 27066880 [c] | 04:56:04.03 | -30:34:01.3 |
| 27066624 [c] | 10:55:59.76 | -77:23:55.1 |
| 27066368 [c] | 10:59:10.00 | -77:23:10.0 |
| 27065088 [c] | 13:00:41.87 | -77:10:22.1 |
| 24402944 [n] | 11:01:55.52 | -34:43:02.1 |
| 27060736 [c] | 16:48:42.53 | -14:15:51.0 |
| 27065600 [c] | 11:10:12.73 | -77:37:09.0 |
| 27065344 [c] | 11:11:39.66 | -76:21:00.3 |

[c] program ID 50641, 20363 (J. Carr, PI)
[n] program ID 30300 (J. Najita, PI)

**Table 3. Measured fluxes of water lines in the spectrum of RNO 90**
(ID program: 50641; AOR 27061760, pointing 1 and 2)

| Wavelength (Microns) | Flux[B] ($10^{-17}$ W m$^{-2}$) | Flux[C] |
|---|---|---|
| 15.16 + 15.17 | < 5.8[B] | 5.7 |
| 15.57 + 15.62 | 18.0 ± 1.0[B] | 16.5 |
| 15.74 | 7.3 ± 0.6[B] | 7.2 |
| 17.10 | 9.0 ± 0.9[B] | 9.5 |
| 17.23 | < 9.4[B] | 9.2 |
| 22.54 | 12.5 ± 1.0[B] | 7.0 |
| 22.62 +22.64 | -- | 7.4 |
| 23.46 + 23.51 | 12.8[B] | 13.0 |
| 26.42 | < 2.0[B] | 2.2 |
| 28.59 | 11.5 ± 0.6[B] | 9.0 |
| 29.14 | < 4.6[B] | 3.9 |
| 29.36 | < 0.8[B] | < 1.0 |
| 30.47 +30.53 | 12.2 ± 0.7[B] | 13.5 |
| 30.87 +30.90 | 10.5 ± 0.8[B] | 10.5 |
| 31.74 +31.77 | < 2.6[B] | 2.5 |
| 32.80 +30.83 | < 3.2[B] | 2.7 |
| 32.92 +32.99+33.01 | 27.4 ± 4.9[B] | 25.0 |

[B]Blevins et al 2016
[C]Cassis (full aperture) Leubutier et al. 2015

**Table 4.** Laboratory wavelengths and integrated absorptivities, ψ, for amino acid mid-IR transitions. Wavelength and flux measurements for bands in the IC 348 combined (averaged) ISM spectrum

**Tryptophan**

| Wavenumber$_{lab}$ (cm$^{-1}$) | Wavelength$_{lab}$ (μm) | Wavelength$_{obs}$ (μm) | ψ (km mol$^{-1}$) | Flux (10$^{-18}$ W m$^{-2}$) | |
|---|---|---|---|---|---|
| 920 | 10.87 | 10.87 | 4.4  | 5.6 | |
| 865 | 11.56 | 11.58 | 10.8 | 4.5 | |
| 744 | 13.43 | 13.45 | 30.6 | 6.4 | |
| 683 | 14.65 | 14.66 | 2.3  | 1.4 | *blend Phe* |
| 627 | 15.96 | 15.96 | 2.7  | 1.1 | |
| 588 | 17.00 | 16.98 | --   | 1.0 | |
| 580 | 17.23 | 17.23 | 5.6  | 4.5 | *blend H$_2$O* |
| 559 | 17.88 | 17.88 | 9.8  | 2.8 | |
| 550 | 18.19 | 18.23 | --   | 2.6 | |
| 528 | 18.92 | 18.93 | --   | 1.2 | |
| 509 | 19.64 | 19.63 | 31.7 | 3.0 | |
| 499 | 20.02 | 20.02 | --   | 0.6 | |
| 457 | 21.88 | 21.88 | 1.9  | 2.0 | |
| 426 | 23.46 | 23.47 | 10.0 | 4.8 | |
| 357 | 28.03 | 28.02 | --   | 1.5 | |
| 349 | 28.65 | 28.70 | 9.6  | 3.0 | |
| 324 | 30.85 | 30.75 | 3.2  | < 3.0 | |

**Tyrosine**

| | | | | | |
|---|---|---|---|---|---|
| 985 | 10.15 | 10.14 | 2.8  | 1.0 | |
| 897 | 11.15 | 11.15 | 2.3  | 0.3 | |
| 877 | 11.40 | 11.45 | 4.6  | 5.0 | |
| 841 | 11.89 | 11.90 | 14.5 | 3.7 | |
| 794 | 12.59 | 12.60 | 12.7 | 3.0 | |
| 740 | 13.50 | 13.51 | 4.9  | 2.5 | |
| 650 | 15.39 | 15.40 | 12.7 | 2.2 | |
| 576 | 17.36 | 17.34 | 12.7 | 6.8 | *blend H$_2$O+Full* |
| 530 | 18.86 | 18.89 | 16.0 | 12.0 | *blend Full* |
| 494 | 20.26 | 20.25 | 2.7  | 2.9 | |
| 434 | 23.05 | 23.01 | 5.5  | 3.0 | |
| 380 | 26.32 | 26.34 | 15.6 | 3.4 | |
| 335 | 29.80 | 29.72 | 1.3  | 2.2 | |

**Phenylalanine**

| | | | | | |
|---|---|---|---|---|---|
| 950 | 10.52 | 10.52 | 7.2 | 5.7 | *blend NH₃* |
| 914 | 10.94 | 10.95 | 1.7 | 2.4 | |
| 779 | 12.84 | 12.86 | 3.6 | 3.0 | |
| 746 | 13.38 | 13.39 | 9.2 | 1.2 | |
| 700 | 14.29 | 14.29 | 19.8 | 6.9 | |
| 526 | 18.98 | 18.98 | 4.0 | 7.2 | |
| 469 | 21.34 | 21.35 | 5.1 | 3.0 | |
| 366 | 27.31 | 27.33 | 54.0 | 10.0 | |

**Isoleucine**

| | | | | | |
|---|---|---|---|---|---|
| 921 | 10.85 | 10.84 | 3.0 | <1.0 | |
| 873 | 11.45 | 11.46 | 3.2 | 2.5 | *blend PAH* |
| 712 | 14.05 | 14.05 | 5.4 | 2.7 | |
| 557 | 17.94 | 17.95 | --- | 3.5 | |
| 538 | 18.59 | 18.58 | 12.9 | 5.0 | |
| 443 | 22.55 | 22.56 | 7.0 | 3.0 | *blend H₂O* |
| 393 | 25.42 | 25.38 | 5.2 | 1.1 | |
| 343 | 29.13 | 29.14 | 4.5 | 3.4 | *blend H₂O* |

**Glycine**

| | | | | | |
|---|---|---|---|---|---|
| 910 | 10.99 | 10.98 | --- | 4.0 | |
| 893 | 11.20 | 10.19 | 10.5 | < 4.0 | *blend PAH* |
| 698 | 14.32 | 14.33 | 6.6 | 13.0 | |
| 608 | 16.45 | 16.45 | 2.1 | 6.0 | |
| 504 | 19.85 | 19.86 | 15.1 | 4.5 | |
| 359 | 27.85 | 27.75 | 14.4 | 3.4 | |

**ACKNOWLEDGEMENTS**

We acknowledge the support of the Instituto de Astrofisica de Canarias and the Ministry of Science and Innovation of Spain via project ESP2015-69020-C2-1-R. We are indebted to Prof. F. Cataldo for his work on characterization of amino acids in the mid-IR . Special acknowledgment to the CASSIS team for the use of their public spectroscopic database.